\documentclass[twocolumn,showpacs,amsfonts,aps,prl,floatfix]{revtex4}

\usepackage{bm}
\usepackage{graphicx}
\usepackage{amsmath}

\def\Np{N_{\mathrm{part}}}
\newcommand{\dNdeta}{dN_\mathrm{ch}/d\eta}

\newcommand{\Tdec}{T_\mathrm{dec}}

\newcommand{\Tchem}{T_\mathrm{chem}}

\def\ve{\varepsilon}
\def\vt#1{v_#1\{2\}}
\def\vf#1{v_#1\{4\}}
\def\et#1{\ve_#1\{2\}}
\def\ef#1{\ve_#1\{4\}}

\def\eq{{\,=\,}}

\usepackage{ulem}
\usepackage{color}

\begin{document}

%%%%%%%%%%%%%%%%%%%%%%%%Front Matter%%%%%%%%%%%%%%%%%%

\title{Hydrodynamic elliptic and triangular flow in Pb--Pb collisions at 
$\bm{\sqrt{s}\eq2.76\,A}$\,TeV}

\author{Zhi Qiu}
\author{Chun Shen}
\author{Ulrich Heinz}
\affiliation{Department of Physics, The Ohio State University,
  Columbus, Ohio 43210-1117, USA}

\begin{abstract}
It is shown that a simultaneous comparison of both elliptic and triangular flow from 
(2+1)-dimensional viscous fluid dynamics with recent measurements in Pb+Pb collisions 
at the Large Hadron Collider (LHC) favors a small specific shear viscosity
$(\eta/s)_\mathrm{QGP}{\,\approx\,}1/(4\pi)$ for the quark-gluon plasma.
Using this viscosity value, the relative magnitude of the elliptic and triangular flow is 
well described with Monte-Carlo Glauber (MC-Glauber) initial conditions while Monte-Carlo 
Kharzeev-Levin-Nardi (MC-KLN) initial conditions require twice as large viscosity to
reproduce the elliptic flow and then underpredict triangular flow by about 30\%.
\end{abstract}


\date{\today}

\maketitle

%%%%%%%%%%%%%%%%%%%%%%%%%%%%%%%%%%%%%%%%%%%%%
%\section{Introduction}
%\label{sec1}
%%%%%%%%%%%%%%%%%%%%%%%%%%%%%%%%%%%%%%%%%%%%%

{\sl 1.\;Introduction.}
Much attention has been given recently to the extraction of the shear viscosity to
entropy density ratio (i.e. the {\it specific shear viscosity} $\eta/s$) of the quark-gluon 
plasma (QGP) from elliptic flow data in relativistic heavy-ion collisions 
\cite{Teaney:2003kp,Lacey:2006pn,Romatschke:2007mq,Luzum:2008cw,%
Luzum:2009sb,Song:2010mg,Aamodt:2010pa,Luzum:2010ag,Lacey:2010ej,%
Bozek:2010wt,Hirano:2010jg,Schenke:2010rr,Schenke:2011tv,Song:2011qa,Shen:2010uy}. 
A major road block in this effort is insufficient knowledge of the initial shape of the 
thermalized fireball created in these collisions, whose initial ellipticity is uncertain by about
 20\% \cite{Hirano:2005xf,Drescher:2006pi,Hirano:2009ah,Heinz:2009cv,Qiu:2011iv}. This 
 induces an ${\cal O}(100\%)$ uncertainty in the value of $(\eta/s)_\mathrm{QGP}$ extracted
from elliptic flow \cite{Luzum:2008cw,Song:2010mg}. After the discovery of
triangular flow in heavy ion collisions at Relativistic Heavy Ion Collider (RHIC) 
\cite{Alver:2010gr,Adare:2011tg,Sorensen:2011fb} and Large Hadron Collider 
(LHC) energies \cite{ALICE:2011vk,CMSflow,Steinberg:2011dj}, followed by the 
confirmation of its collective hydrodynamic nature \cite{Alver:2010gr,Alver:2010dn,%
Petersen:2010cw,Qin:2010pf,Luzum:2010sp,Xu:2011fe,Luzum:2011mm} and the 
realization that shear viscosity suppresses higher order harmonic flow coefficients 
more strongly than elliptic flow \cite{Alver:2010dn,Schenke:2010rr,Schenke:2011tv,%
Chaudhuri:2011qm,Schenke:2011bn}, it was recently suggested \cite{Lacey:2010hw,%
Adare:2011tg,ALICE:2011vk,Shen:2011zc,Qiu:2011fi} that a combined analysis of the elliptic
and triangular flow coefficients $v_2$ and $v_3$ could yield a more precise value
for the QGP shear viscosity and thereby reduce or eliminate the model uncertainty
in the initial deformation of the QGP fireball and its event-by-event fluctuations.
This Letter presents such an analysis.

Our study is based on a (2+1)-dimensional viscous hydrodynamic model with 
longitudinal boost-invariance, describing numerically the transverse evolution 
of the heavy-ion collision fireball near midrapidity. As in past work 
\cite{Luzum:2008cw,Hirano:2009ah,Luzum:2009sb,Heinz:2009cv,Qiu:2011iv,Song:2010mg} 
we explore two different types of fluctuating initial conditions for the entropy and energy 
density profiles, generated from Monte Carlo versions of the Glauber and KLN models 
(see \cite{Hirano:2009ah} and references therein for details of the implementation used
here). 

\begin{figure*}
  \begin{tabular}{cc}
  \includegraphics[width=0.4\linewidth]{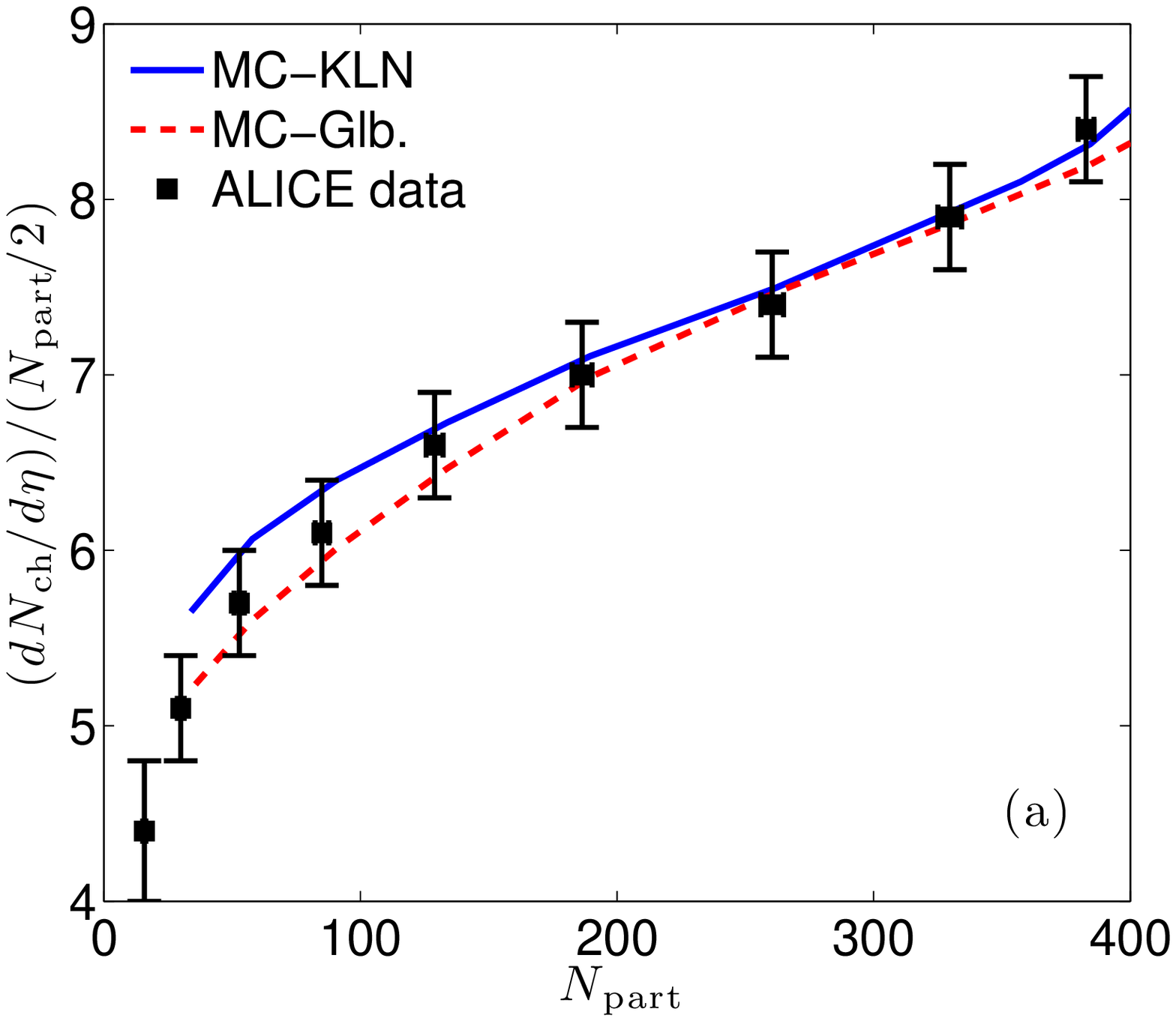} &
  \includegraphics[width=0.4\linewidth]{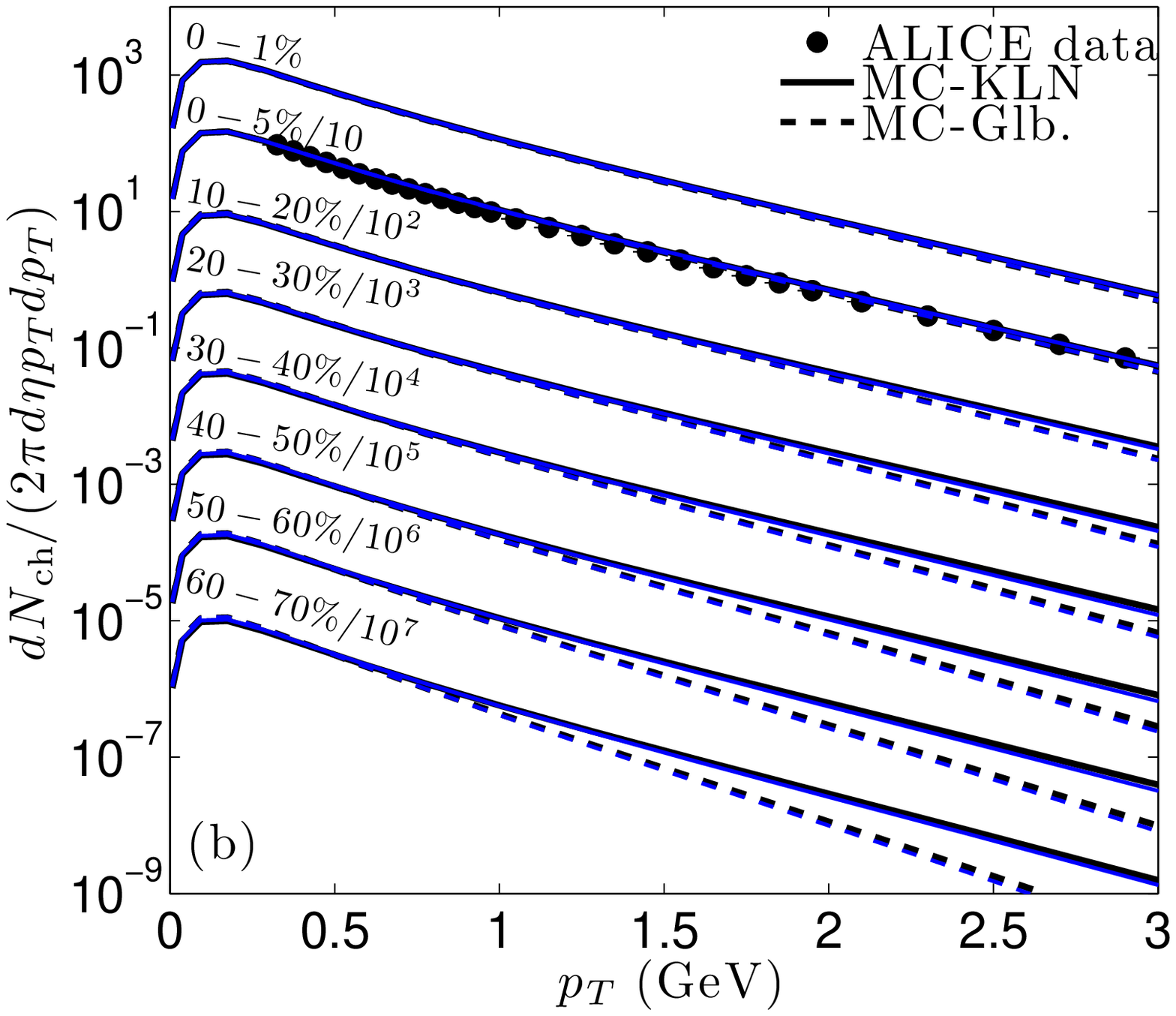}
  \end{tabular}
  \caption{(a) (Color online) Centrality dependence of charged particle multiplicity density 
    as a function of $\Np$ from the MC-Glauber (dashed) and MC-KLN (solid) models, 
    compared with ALICE measurements \cite{Aamodt:2010cz} for 2.76\,$A$\,TeV Pb-Pb 
    collisions. 
    (b) Charged particle $p_T$-spectra from the MC-Glauber and MC-KLN models for 
    different centralities. The most central ($0{-}5\%$) results are compared with 
    ALICE data \cite{Aamodt:2010jd}.}
  \label{fig:1}
\end{figure*}

The MC-KLN calculations were done using a Monte-Carlo sample of initial state profiles 
with identical properties as those used in \cite{Shen:2011eg} for the calculation
of transverse momentum spectra and elliptic flow in 2.76\,$A$\,TeV Pb-Pb collisions at the
LHC. For the $x$ dependence of the gluon structure function in the MC-KLN model we
used the power $\lambda\eq0.28$ \cite{Hirano:2009ah}; the normalization factor for the
initial entropy density was fixed by hand to reproduce the measured charged hadron
multiplicity density $\dNdeta$ for the 5\% most central collisions \cite{Aamodt:2010pb};
the measured dependence of $\dNdeta$ on collision centrality \cite{Aamodt:2010cz}
is then automatically reproduced reasonably well by the model \cite{Shen:2011eg} (see 
Fig.~\ref{fig:1}(a)). MC-KLN runs were done with $\eta/s\eq0.2$ which, for this type of 
initial conditions, was shown to yield a good overall description of the measured 
transverse momentum spectra and elliptic flow in 200\,$A$\,GeV Au-Au collisions 
at RHIC \cite{Shen:2011eg} and gave an impressively accurate prediction for the 
unidentified and identified charged hadron spectra and elliptic flows in 2.76\,$A$\,TeV 
Pb-Pb collisions at the LHC \cite{Shen:2011eg,Heinz:2011kt}. 

For the MC-Glauber runs we generated a new set of initial configurations that differ
from those used for 200\,$A$\,GeV Au-Au collisions in \cite{Song:2010mg} by 
the wounded nucleon to binary collision ratio. Taking the initial entropy density
$s(\bm{r}_\perp;b)\eq\kappa\bigl(\frac{1{-}x}{2}n_{_\mathrm{WN}}(\bm{r}_\perp;b) + x  
n_{_\mathrm{BC}}(\bm{r}_\perp;b)\bigr)$, we determine $\kappa$ and $x$ by a two-parameter
fit to the ALICE data \cite{Aamodt:2010cz} shown in Fig.~\ref{fig:1}(a). Due to viscous
entropy production during the hydrodynamic evolution, which itself depends on collision 
centrality, the fit value for $x$ depends on the assumed shear viscosity. For MC-Glauber 
initial conditions we  took $\eta/s\eq0.08$ since this value was shown in 
\cite{Schenke:2011tv, Schenke:2011bn,ALICE:2011vk} to provide a reasonable 
description of the charged hadron $v_2(p_T)$ and $v_3(p_T)$ data measured by the 
ALICE experiment; this results in $x\eq0.118$ for Pb-Pb collisions at the LHC. Both the 
MC-Glauber and MC-KLN initial conditions are hydrodynamically evolved with equation 
of state (EOS) s95p-PCE \cite{Shen:2010uy} which matches numerical results from lattice 
QCD at high temperatures to a hadron resonance gas at low temperatures 
\cite{Huovinen:2009yb} and implements chemical freeze-out at $\Tchem\eq165$\,MeV. 
The hydrodynamic output  is converted to final hadron distributions along an isothermal 
decoupling surface of  temperature $\Tdec\eq120$\,MeV, using the Cooper-Frye prescription.

%===================   Fig. 1 ====================
%=======================================

In \cite{Qiu:2011iv} we showed that, due to similar fluctuation mechanisms, the
MC-KLN and MC-Glauber models generate similar third-order eccentricities 
$\ve_3$ whereas the ellipticity $\ve_2$, which is mostly controlled by collision geometry,
is about 20\% larger in the MC-KLN model. Event-by-event ideal \cite{Qiu:2011iv}
and viscous \cite{Qiu:2011fi,Qiu:future1} hydrodynamic simulations with both realistically
fluctuating \cite{Qiu:2011iv,Qiu:future1} and doubly-deformed Gaussian initial
conditions \cite{Qiu:2011fi} (with simultaneously non-zero $\ve_2$ and $\ve_3$ 
eccentricities) have shown that the hydrodynamic conversion efficiencies for translating 
initial spatial eccentricities into final flow anisotropies  \cite{Qin:2010pf,%
Teaney:2010vd,Shuryak:2009cy}, although different for $v_2/ \ve_2$ and $v_3/\ve_3$,
are very similar in the MC-KLN and MC-Glauber models. The similarities in $\ve_3$ and
differences in $\ve_2$ between these models should thus straightforwardly reflect 
themselves in analogous differences in $v_2$ and $v_3$ \cite{Shen:2011zc,Qiu:2011fi}, 
allowing for an experimental distinction between the models.

Event-by-event viscous hydrodynamic simulations with full inclusion of unstable 
resonance decays are at present numerically too costly for systematic flow studies
over a range of viscosities, collision energies, centralities, and collision systems. A 
recent event-by-event study by Schenke {\it et al.} \cite{Schenke:2011bn} with a 
restricted set of resonances showed that, compared to a full calculation, $v_2$ ($v_3$)
was overpredicted by 10-15\% (25-30\%). This is larger than the difference in
these observables seen \cite{Qiu:2011iv,Qiu:2011fi,Qiu:future1} between 
event-by-event calculations and ``single-shot hydrodynamics" where the fluctuating
initial conditions are averaged (after rotating each event by its second or third order
participant plane angle \cite{Qiu:2011iv,Qiu:2011fi} to align the respective eccentricities)
{\it before} the hydrodynamic evolution instead of afterwards. For this reason we here
use the single-shot approach, but include the full cascade of resonance decays in
the final state. Our approach here differs from that in \cite{Alver:2010dn} by replacing 
the singly-deformed Gaussian parametrization of the initial density used there by
the ensemble-average of realistically fluctuating, non-Gaussian initial profiles, and 
from \cite{Alver:2010dn} and \cite{Schenke:2010rr,Schenke:2011tv,Schenke:2011bn}
by employing a more realistic EOS that accounts for the important effects of chemical
non-equilibrium hadronic evolution on the elliptic flow $v_2$ \cite{Hirano:2002ds}.
In \cite{ALICE:2011vk} it was shown that, with the approach used in \cite{Alver:2010dn},
MC-KLN initial conditions with $\eta/s\eq0.16$ cannot describe the $p_T$-integrated
$v_3$ measured in 2.76\,$A$\,TeV Pb-Pb collisions, whereas the MC-Glauber based
event-by-event calculations (with $\eta/s\eq0.08$) of Schenke {\it et al.}
\cite{Schenke:2011tv} appear to describe $v_3(p_T)$ at selected centralities reasonably 
well.

%%%%%%%%%%%%%%%%%%%%%%%%%%%%%%%%%%%%%%%%%%%%%
%\section{Multiplicity and Spectra}
%\label{sec2}
%%%%%%%%%%%%%%%%%%%%%%%%%%%%%%%%%%%%%%%%%%%%%

%================  Fig. 2  =======================
\begin{figure}[b]
  \includegraphics[width=0.9\linewidth]{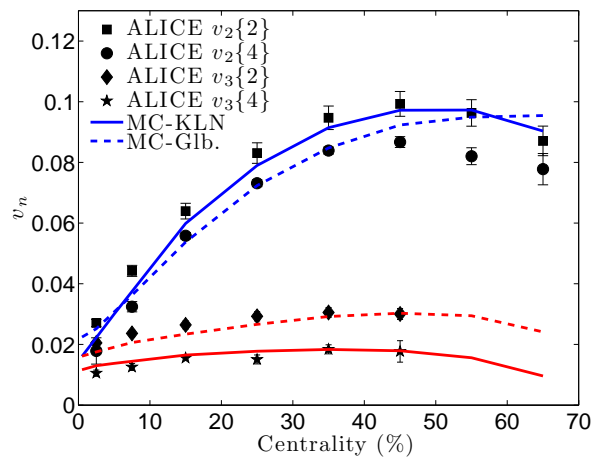}
  \caption{$v_2$ and $v_3$ vs. centrality, compared with ALICE $\vt{2}$, $\vf{2}$, 
         $\vt{3}$, and $\vf{3}$ data for 2.76\,$A$\,TeV Pb+Pb \cite{ALICE:2011vk}.}
  \label{fig:2}
\end{figure}
%=============================================

%=================  Fig. 3  ======================
\begin{figure*}
  \begin{tabular}{cc}
  \includegraphics[width=0.4\linewidth]{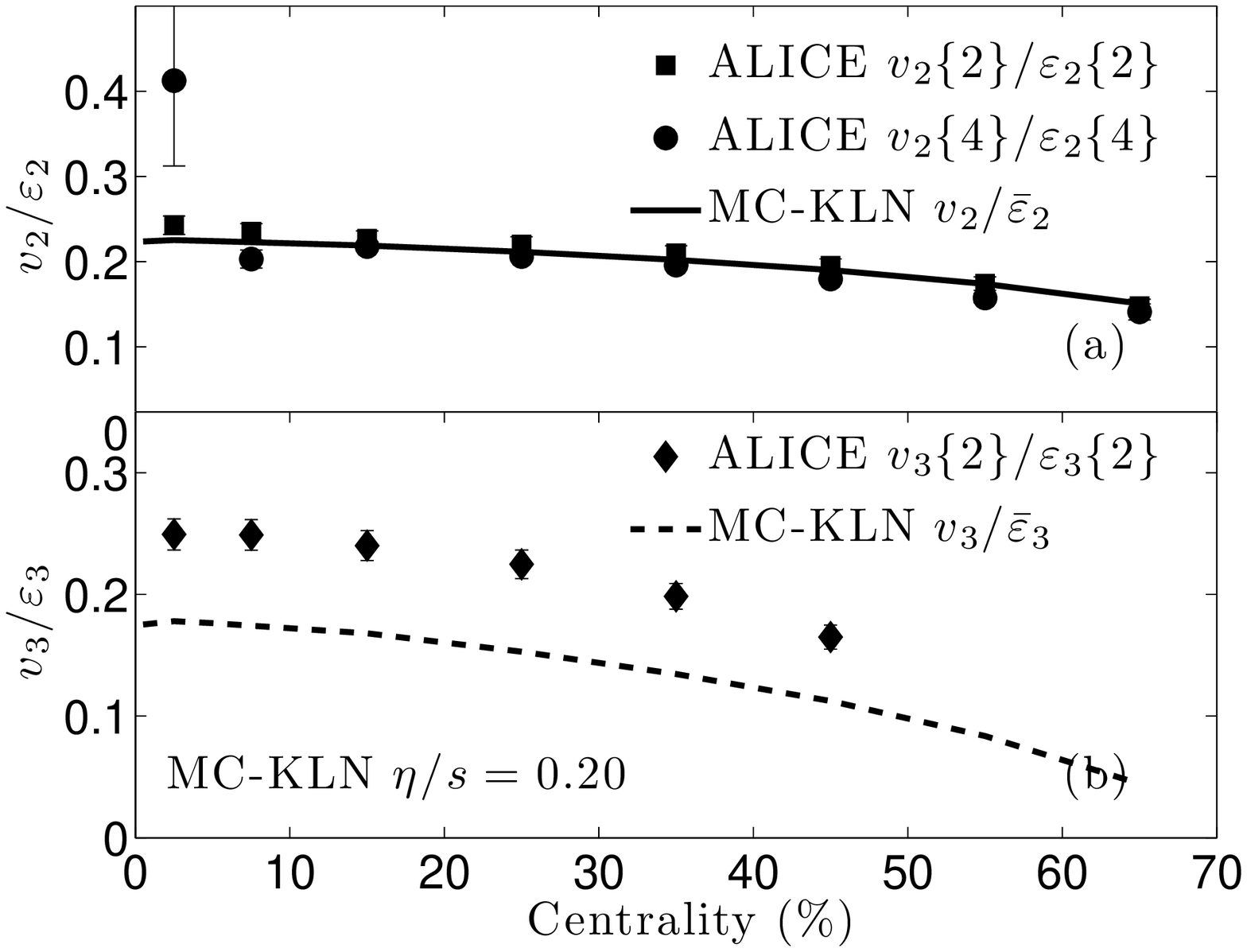} &
  \includegraphics[width=0.4\linewidth]{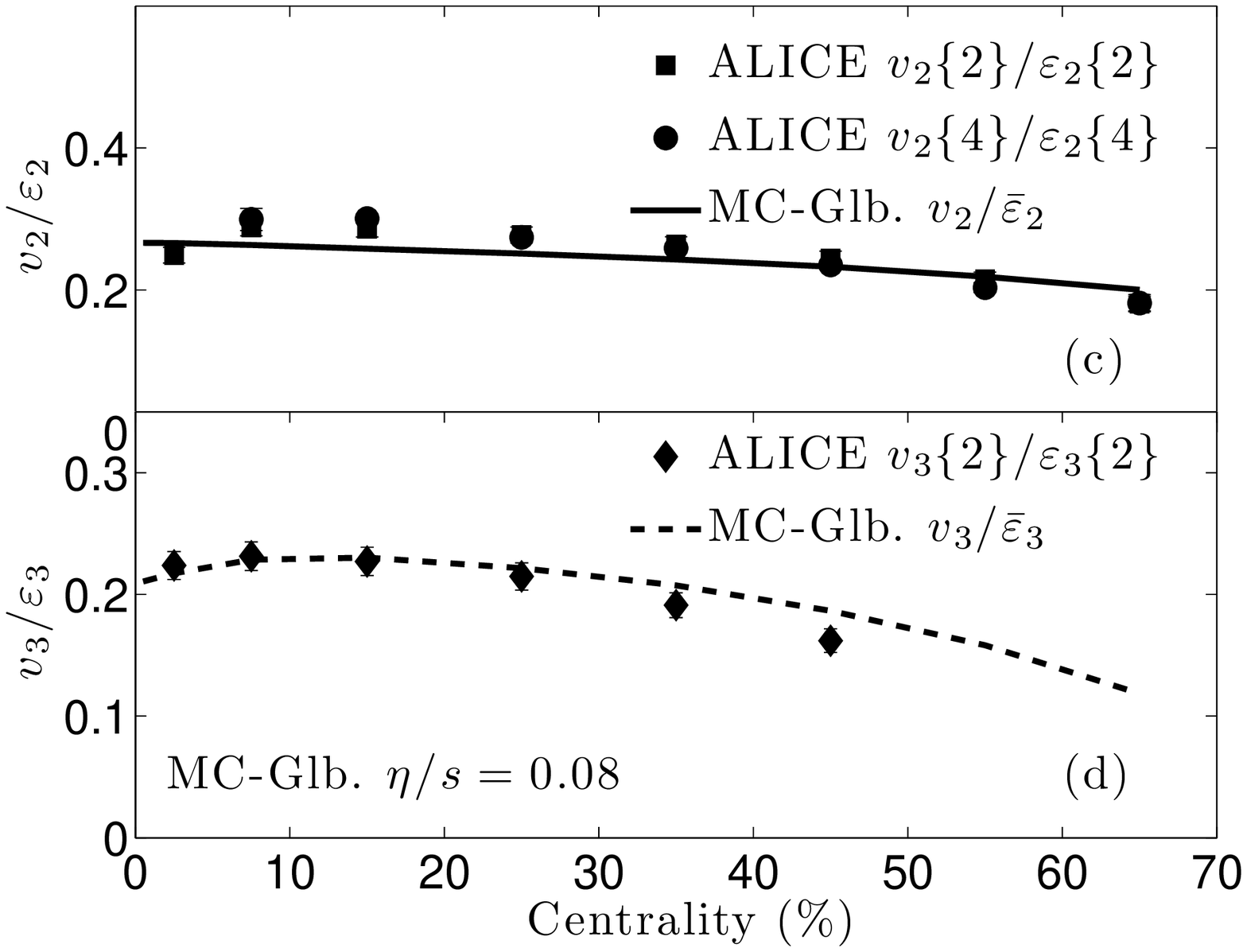}
  \end{tabular}
  \caption{Eccentricity-scaled, $p_T$-integrated $v_{2,3}$ for the hydrodynamically evolved 
    MC-KLN (a,b) and MC-Glauber (c,d) models, compared with ALICE $v_{2,3}$ data 
    for 2.76\,$A$\,TeV Pb-Pb collisions \cite{ALICE:2011vk} scaled by their corresponding
    eccentricities (see text). 
    \label{fig:3}}
\end{figure*}
%=============================================

{\sl 2.\;Transverse momentum spectra.}
Figure\;\ref{fig:1}(b) shows the charged hadron $p_T$-spectra for 2.76\,$A$\,TeV Pb-Pb 
collisions at different centralities, for both MC-Glauber ($\eta/s\eq0.08$) and MC-KLN 
($\eta/s\eq0.2$) initial conditions. For the most central ($0{-}5\%$) collisions the spectra
from both models agree well with published ALICE data. In more peripheral collisions
the MC-KLN spectra are harder than those from MC-Glauber initial conditions. This is
a consequence of larger radial flow caused by larger transverse viscous pressure
gradients in the MC-KLN case where the fluid is taken to have 2.5 times larger shear
viscosity than for the MC-Glauber simulations, in order to obtain the same elliptic flow
\cite{Luzum:2008cw,Song:2010mg}. In peripheral collisions these viscous effects
are stronger than in more central collisions where the fireball is
larger \cite{Song:2008si}. As shown in \cite{Holopainen:2010gz,Qiu:2011iv},
event-by-event evolution of fluctuating initial conditions generates, for small values 
of $\eta/s$, flatter hadron spectra than single-shot hydrodynamics, especially in 
peripheral collisions, due to stronger radial flow driven by hot spots in the fluctuating 
initial states. Proper event-by-event evolution of the latter is therefore expected 
to reduce the difference between the MC-Glauber and MC-KLN curves in 
Fig.~\ref{fig:1}(b) since this effect is relatively strong for $\eta/s\eq0.08$ (MC-Glauber)
\cite{Qiu:2011iv} but almost absent for $\eta/s\eq0.2$ (MC-KLN) \cite{Qiu:future1}.

%%%%%%%%%%%%%%%%%%%%%%%%%%%%%%%%%%%%%%%%%%%%%
%\section{$p_T$ integrated flow}
%\label{sec3}
%%%%%%%%%%%%%%%%%%%%%%%%%%%%%%%%%%%%%%%%%%%%%

{\sl 3.\;$p_T$-integrated elliptic and triangular flow.}
In Figure\;\ref{fig:2} we compare our $p_T$-integrated $v_2$ and $v_3$ as functions 
of centrality with ALICE $\vt{2}$, $\vf{2}$, $\vt{3}$, and $\vf{3}$ data, extracted from 
2- and 4-particle correlations \cite{ALICE:2011vk}. For both models, $v_{2,3}$ from 
averaged smooth initial conditions lie between the experimental $\vt{{2,3}}$ and $\vf{{2,3}}$
values. This is consistent with the theoretical expectation \cite{Miller:2003kd,Voloshin:2007pc}
that $\vt{n}$ ($\vf{n}$) is shifted up (down) relative to the average flow by event-by-event 
flow fluctuations and was also found elsewhere \cite{Song:2010mg, Schenke:2011tv,%
Luzum:2010ag}. Upon closer inspection, however, and recalling that ideal single-shot hydrodynamics with smooth initial condition was shown \cite{Qiu:2011iv} to generate $v_2$ 
similar to $\vt{2}$ from the corresponding event-by-event evolution, it seems that the 
MC-KLN is favored since it produces $v_2$ results closer to the $\vt{2}$ data. Unfortunately,
a similar argument using $v_3$ can be held against the MC-KLN model. To eliminate the 
interpretation difficulties associated with a comparison of average flows from single-shot 
evolution of averaged initial conditions with data affected irreducibly by naturally existing 
event-by-event fluctuations, we proceed to a comparison of eccentricity-scaled flow 
coefficients.

Assuming linear response of $v_{2,3}$ to their respective eccentricities $\ve_{2,3}$
(which was found to hold with reasonable accuracy for $v_2$ and $v_3$ but not for
higher order anisotropic flows \cite{Qiu:2011iv}), we follow \cite{Bhalerao:2006tp}
and scale the flow $v_{2,3}$ from single-shot hydrodynamics by the eccentricity
$\bar{\ve}_{2,3}$ of the ensemble-averaged smooth initial energy density,
while scaling the experimental  $\vt{{2,3}}$ and $\vf{{2,3}}$ data by the corresponding 
fluctuating eccentricity measures $\et{{2,3}}$ and $\ef{{2,3}}$, respectively, calculated 
from the corresponding models. In \cite{Qiu:future1} we justify this procedure 
for $\vt{{2,3}}$ and $\vf{2}$ and also show that it fails for $\vf{3}/\ef{3}$ since this ratio 
is found to differ strongly from $v_3/\bar{\ve}_3$.

%===================  Fig 4  ====================
\begin{figure*}
  \begin{tabular}{cc}
  \includegraphics[width=0.4\linewidth]{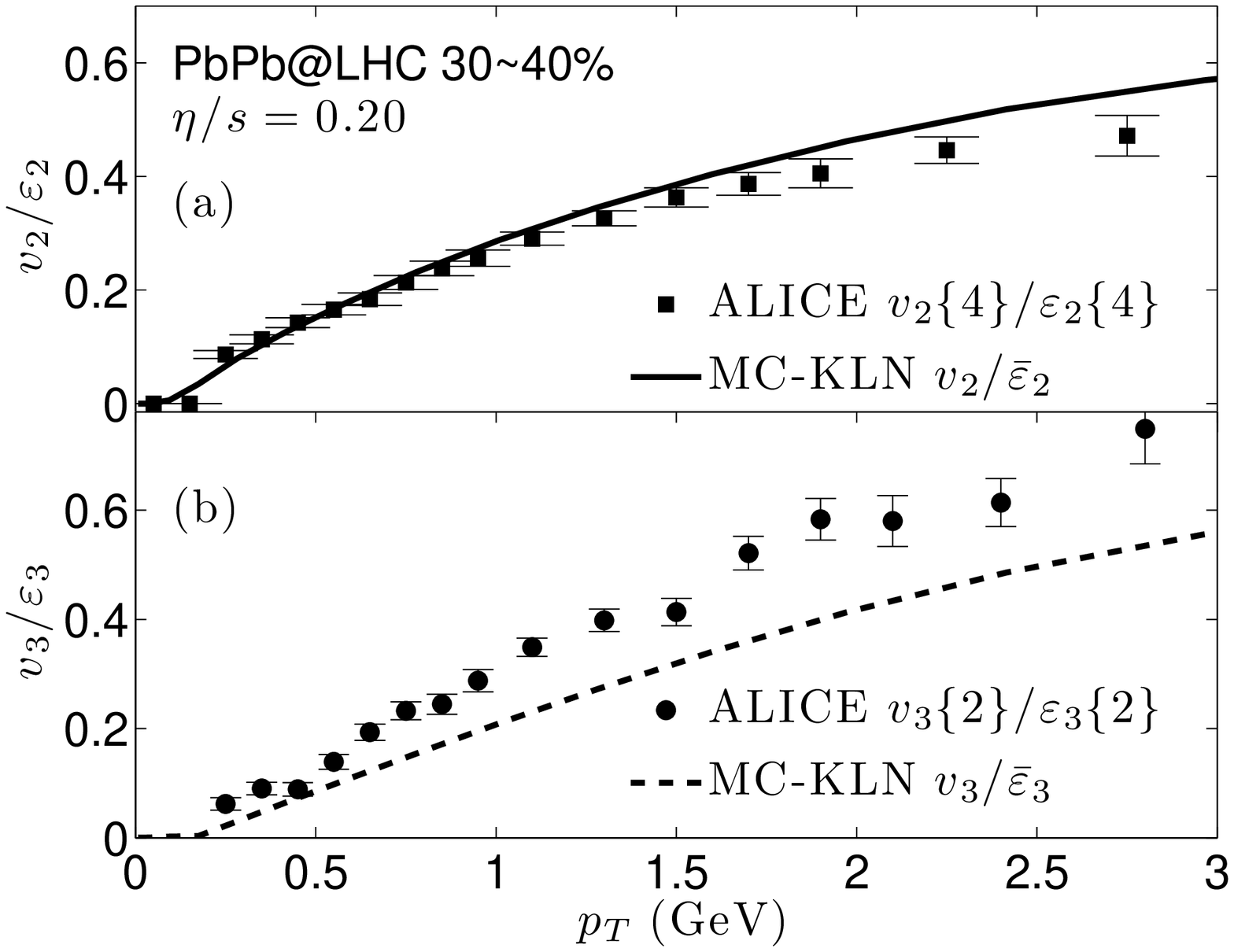} &
  \includegraphics[width=0.41\linewidth]{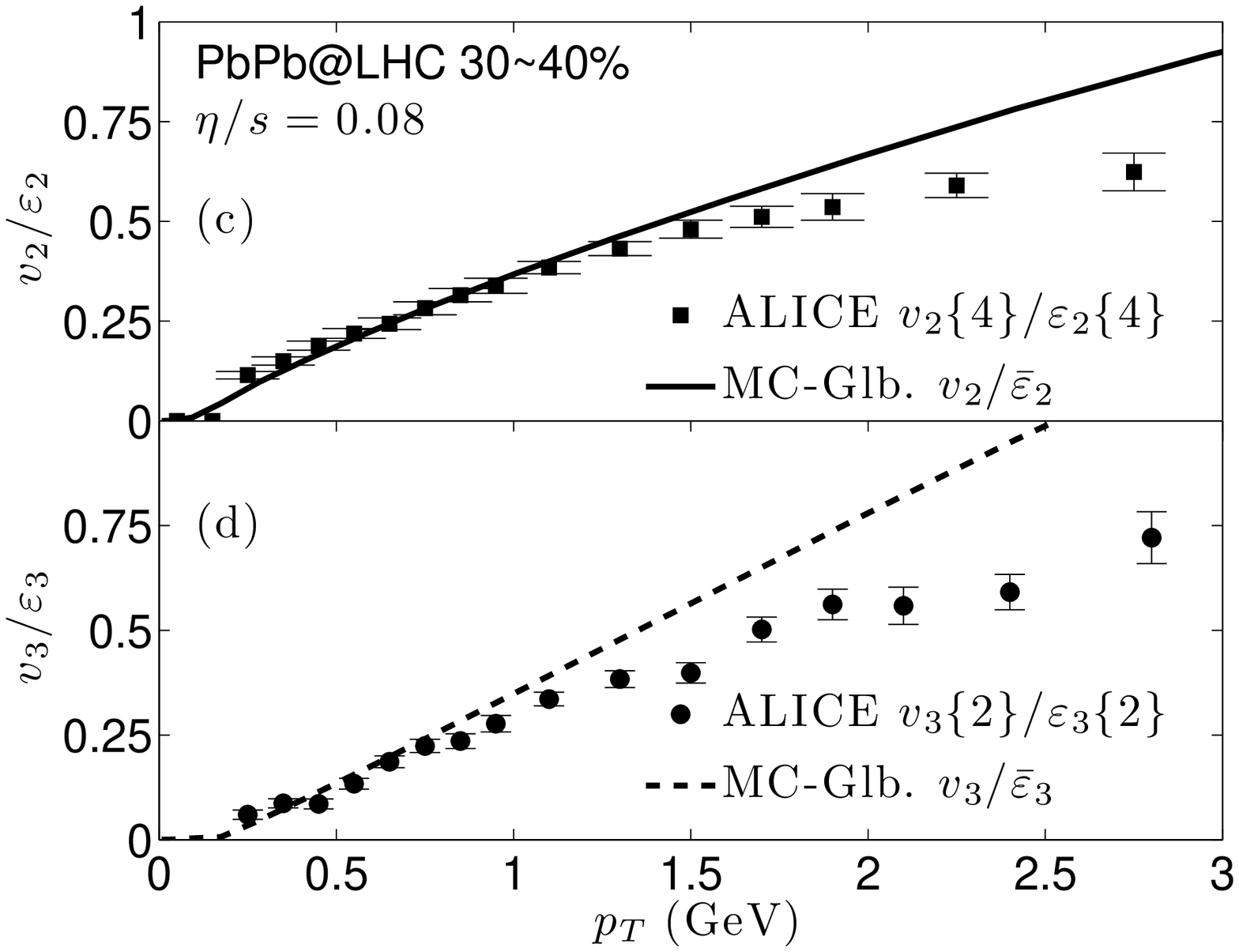}
  \end{tabular}
  \caption{Eccentricity-scaled, $p_T$-differential elliptic and triangular flow for 
     2.76\,$A$\,TeV Pb-Pb collisions from viscous hydrodynamics with MC-KLN (a,b) 
     and MC-Glauber (c,d) initial conditions. The ALICE data \cite{ALICE:2011vk} are 
     scaled according to their corresponding eccentricities, see text.}
  \label{fig:4}
\end{figure*}
%=============================================

The eccentricity-scaled elliptic and triangular flow coefficients for the MC-KLN 
and MC-Glauber models are shown in Figs.~\ref{fig:3}(a,b) and \ref{fig:3}(c,d),
respectively, and compared with the corresponding data from ALICE. The first thing
to note is the impressively accurate agreement between the experimentally
measured $\vt{2}/\et{2}$ and $\vf{2}/\ef{2}$, showing that for elliptic flow the idea
of scaling ``each flow with its own eccentricity" \cite{Bhalerao:2006tp} works very well.
The same is not true for $\vt{3}/\et{3}$ and $\vf{3}/\ef{3}$ for which the experimental
do not at all agree (not shown), nor are they expected to \cite{Qiu:future1}. Secondly,
both $\vt{2}/\et{2}$ and $\vf{2}/\ef{2}$ measured by ALICE agree well with the viscous
hydrodynamic calculations, for both the MC-Glauber and MC-KLN models, confirming 
that for each model the correct value of $\eta/s$ has been used as far as elliptic flow
is concerned.

The bottom panels in Fig.~\ref{fig:3} show the triangular flow $v_3$. Clearly, with the
viscosities needed to reproduce $v_2$, the MC-KLN model badly disagrees with
the experimental data. The measured triangular flow is too big to accommodate
a specific shear viscosity as large as 0.2. Within the present approach, the only 
possibility to avoid this conclusion is that somehow the MC-Glauber and MC-KLN 
models both underpredict the initial third-order eccentricity $\ve_3$ by about 50\%.
With MC-Glauber initial conditions and $\eta/s\eq0.08$, on the other hand, the 
ALICE data agree well with viscous hydrodynamics, even if the measured centrality 
dependence of $\vt{3}/\et{3}$ is slightly steeper than the calculated one. 
 
Summarizing Fig.~\ref{fig:3}, the ALICE data for the $p_T$-integrated elliptic and 
triangular data strongly favor MC-Glauber initial conditions and, by implication, a 
small value of  $\eta/s{\,\simeq\,}0.08$ for the specific QGP shear viscosity.

%%%%%%%%%%%%%%%%%%%%%%%%%%%%%%%%%%%%%%%%%%%%%
%\section{$p_T$ differential flow}
%\label{sec4}
%%%%%%%%%%%%%%%%%%%%%%%%%%%%%%%%%%%%%%%%%%%%%

{\sl 4.\;$p_T$-differential elliptic and triangular flow.}
We close this Letter by cross-checking, at one collision centrality ($30{-}40\%$) where
$v_3(p_T)$ data are available \cite{ALICE:2011vk}, the $p_T$-differential anisotropic 
flows. The corresponding comparison between data and theory is shown in Fig.~\ref{fig:4};
as in Fig.~\ref{fig:3} we compare the eccentricity-scaled flows, plotting $v_{2,3}/\bar{\ve}_{2,3}$
for the models and $\vf{2}/\ef{2}$ ($\vt{3}/\et{3}$) for the elliptic (triangular) flow data.
As seen in the upper panels, both initial state models describe the measured
elliptic flow well up to $p_T{\,\sim\,}1{-}1.5$\,GeV/$c$; at larger $p_T$, they overpredict
$v_2(p_T)$ for charged particles -- a problem noticed before \cite{Song:2011qa,Shen:2011eg}
and possibly related to an imperfect model description of the measured final chemical
composition \cite{Heinz:2011kt}. The disagreement at larger $p_T$ is worse for
MC-Glauber initial conditions; this is likely related to our earlier observation in 
Fig.~\ref{fig:1}(b) that our MC-Glauber $p_T$-spectra are steeper than the MC-KLN ones
in peripheral collisions -- an artifact of our single-shot approach and possibly remedied
by a proper event-by-event hydrodynamical simulation.

Figure~\ref{fig:4}(b) shows again the disagreement between theory and
experiment for triangular flow when we use MC-KLN initial conditions: the model
strongly underpredicts the data at all $p_T$, i.e. it gives the wrong slope for
$v_3(p_T)$. With MC-Glauber initial conditions and correspondingly lower
shear viscosity $\eta/s\eq0.08$ (Fig.~\ref{fig:4}(d)), the measured $v_3(p_T)$ is 
well described up to $p_T{\,\sim\,}1$\,GeV/$c$ but overpredicted at larger $p_T$.
Again, the latter can be at least partially attributed to the fact that MC-Glauber 
$p_T$-spectrum from our single-shot hydrodynamic approach is too steep at this 
collision centrality, which can in future studies be corrected by performing the
hydrodynamic evolution properly event by event.  
 
%%%%%%%%%%%%%%%%%%%%%%%%%%%%%%%%%%%%%%%%%%%%%
%\section{Summary}
%\label{sec5}
%%%%%%%%%%%%%%%%%%%%%%%%%%%%%%%%%%%%%%%%%%%%%

{\sl 5.\;Summary.} Using a well-calibrated single-shot viscous hydrodynamic approach 
without hadronic cascade afterburner but properly implementing hadronic chemical 
freeze-out at $\Tchem{\,\approx\,}165$\,MeV and including a full set of resonance 
decays, we have shown that a combined analysis of the ALICE data for elliptic 
and triangular flow from 2.76\,$A$\,TeV Pb-Pb collisions leads to a strong preference
for initial conditions from the Monte-Carlo Glauber model, combined with a low
value for the QGP shear viscosity $\eta/s{\,\simeq\,}0.08$, and disfavors the considerably
larger viscosities of $\eta/s{\,\sim\,}0.2$ that are required to reproduce the measured
elliptic flow when assuming the more eccentric Monte-Carlo KLN initial profiles. Final
confirmation of these conclusions will require a proper event-by-event evolution of
the fluctuating initial density profiles and coupling viscous hydrodynamics to a
microscopic description of the dilute late hadronic stage where viscous hydrodynamics
breaks down \cite{Song:2010aq}, and a similar analysis of recently published PHENIX
data at lower RHIC energies \cite{Adare:2011tg}. Given the large magnitude of the 
underprediction  $v_3$ in the MC-KLN model observed here we doubt, however, that
such more sophisticated approaches will be able to reverse the conclusions drawn here.

{\sl Acknowledgments:} 
We thank R. Snellings for providing us with tables of the experimental data from the ALICE experiment and for helpful discussions. This work was supported by the U.S.\ Department of Energy under Grants No.~\rm{DE-SC0004286} and (within the framework of the JET 
Collaboration) \rm{DE-SC0004104}. 

%%%%%%%%%%%%%%% References %%%%%%%%%%%%%%%%%%%%%%%%%%%

%\bibliographystyle{h-physrev3}
%\bibliography{references}

\begin{thebibliography}{99}

\bibitem{Teaney:2003kp}
  D.~Teaney,
  %``The Effects of viscosity on spectra, elliptic flow, and HBT radii,''
  Phys.\ Rev.\  C {\bf 68}, 034913 (2003).
  %[nucl-th/0301099].
  
\bibitem{Lacey:2006pn}
  R.~A.~Lacey and A.~Taranenko,
  %``What do elliptic flow measurements tell us about the matter created in
  %  the little bang at RHIC?,''
  PoS {\bf CFRNC2006}, 021 (2006);
  %[nucl-ex/0610029].
%\bibitem{Lacey:2006bc}
  R.~A.~Lacey {\it et al.},
  %``Has the QCD Critical Point been Signaled by Observations at RHIC?,''
  Phys.\ Rev.\ Lett.\  {\bf 98}, 092301 (2007);
  %[nucl-ex/0609025].
%\bibitem{Adare:2006nq}
  A.~Adare {\it et al.},
  %(PHENIX Collaboration),
  %``Energy Loss and Flow of Heavy Quarks in Au+Au Collisions at
  %  s(NN)**(1/2) = 200-GeV,''
  Phys.\ Rev.\ Lett.\  {\bf 98}, 172301 (2007);
  %[nucl-ex/0611018].
%\bibitem{Drescher:2007cd}
  H.-J.~Drescher, A.~Dumitru, C.~Gombeaud, and J.-Y. Ollitrault,
  %``The Centrality dependence of elliptic flow, the hydrodynamic limit,
  %  and the viscosity of hot QCD,''
  Phys.\ Rev.\  C {\bf 76}, 024905 (2007).
  %[arXiv:0704.3553 [nucl-th]].
%\bibitem{Dusling:2007gi}
  K.~Dusling and D.~Teaney,
  %``Simulating elliptic flow with viscous hydrodynamics,''
  Phys.\ Rev.\  C {\bf 77}, 034905 (2008);
  %[arXiv:0710.5932 [nucl-th]].
  %%CITATION = PHRVA,C77,034905;%%
%\bibitem{Xu:2007jv}
  Z.~Xu, C.~Greiner, and H.~St\"ocker,
  %``PQCD calculations of elliptic flow and shear viscosity at RHIC,''
  Phys.\ Rev.\ Lett.\  {\bf 101}, 082302 (2008);
  %[arXiv:0711.0961 [nucl-th]].
%\bibitem{Molnar:2008xj}
  D.~Molnar and P.~Huovinen,
  %``Dissipative effects from transport and viscous hydrodynamics,''
  J.\ Phys.\ G {\bf 35}, 104125 (2008);
  %[arXiv:0806.1367 [nucl-th]].
  %%CITATION = JPHGB,G35,104125;%%
%\bibitem{Lacey:2009xx}
  R.~A.~Lacey, A.~Taranenko and R.~Wei,
  %``Is the quark gluon plasma produced in RHIC collisions strongly coupled?,''
  in {\it Proc. 25th Winter Workshop on Nuclear Dynamics},
  W. Bauer, R. Bellwied, and J.W. Harris (eds.),
  (EP Systema, Budapest, 2009) p. 73 [arXiv:0905.4368 [nucl-ex]];
  %%CITATION = ARXIV:0905.4368;%%
%\bibitem{Dusling:2009df}
  K.~Dusling, G.~D.~Moore, and D.~Teaney,
  %``Radiative energy loss and v2 spectra for viscous hydrodynamics,''
  Phys.\ Rev.\  C {\bf 81}, 034907 (2010);
  %[arXiv:0909.0754 [nucl-th]].
  %%CITATION = PHRVA,C81,034907;%%
%\bibitem{Chaudhuri:2009hj}
  A.~K.~Chaudhuri,
  %``Centrality dependence of elliptic flow and QGP viscosity,''
  J.\ Phys.\ G {\bf 37}, 075011 (2010);
  %[arXiv:0910.0979 [nucl-th]].
%\bibitem{Lacey:2010fe}
  R.~A.~Lacey {\it et al.},
  %``Azimuthal anisotropy: transition from hydrodynamic flow to jet
  %  suppression,''
  Phys.\ Rev.\  C {\bf 82}, 034910 (2010).
  %[arXiv:1005.4979 [nucl-ex]].

\bibitem{Romatschke:2007mq}
  P.~Romatschke and U.~Romatschke,
  %``Viscosity Information from Relativistic Nuclear Collisions: How Perfect is the Fluid Observed at RHIC?,''
  Phys.\ Rev.\ Lett.\  {\bf 99}, 172301 (2007).
  %[arXiv:0706.1522 [nucl-th]].

\bibitem{Luzum:2008cw}
  M.~Luzum and P.~Romatschke,
  %``Conformal Relativistic Viscous Hydrodynamics: Applications to RHIC
  %  results at s(NN)**(1/2) = 200-GeV,''
  Phys.\ Rev.\  C {\bf 78}, 034915 (2008).
  %[Erratum {\it ibid.} C {\bf 79}, 039903(E) (2009)].
  %[arXiv:0804.4015 [nucl-th]].

\bibitem{Luzum:2009sb}
  M.~Luzum and P.~Romatschke,
  %``Viscous Hydrodynamic Predictions for Nuclear Collisions at the LHC,''
  Phys.\ Rev.\ Lett.\  {\bf 103}, 262302 (2009).
  %[arXiv:0901.4588 [nucl-th]].

\bibitem{Song:2010mg}
  H.~Song, S.~A.~Bass, U.~Heinz, T.~Hirano and C.~Shen,
  %``200 A GeV Au+Au collisions serve a nearly perfect quark-gluon liquid,''
  Phys.\ Rev.\ Lett.\ {\bf 106}, 192301 (2011);
  %%CITATION = ARXIV:1011.2783;%%
%\bibitem{Song:2011hk}
  %H.~Song, S.~A.~Bass, U.~Heinz, T.~Hirano and C.~Shen,
  %``Hadron spectra and elliptic flow for 200 A GeV Au+Au collisions from
  %viscous hydrodynamics coupled to a Boltzmann cascade,''
  and Phys.\ Rev.\ C {\bf 83}, 054910 (2011).
  %[arXiv:1101.4638].
  %%CITATION = ARXIV:1101.4638;%%

\bibitem{Aamodt:2010pa}
  K.~Aamodt {\it et al.} (ALICE Collaboration),
  %``Elliptic flow of charged particles in Pb-Pb collisions at 2.76 TeV,''
  Phys.\ Rev.\ Lett.\  {\bf 105}, 252302 (2011).
  %arXiv:1011.3914 [nucl-ex].
  %%CITATION = ARXIV:1011.3914;%%

\bibitem{Luzum:2010ag}
  M.~Luzum,
  %``Elliptic flow at energies available at the CERN Large Hadron Collider:
  %Comparing heavy-ion data to viscous hydrodynamic predictions,''
  Phys.\ Rev.\  C {\bf 83}, 044911 (2011).
  %[arXiv:1011.5173 [nucl-th]].
  %%CITATION = PHRVA,C83,044911;%%

\bibitem{Lacey:2010ej}
  R.~A.~Lacey, A.~Taranenko, N.~N.~Ajitanand and J.~M.~Alexander,
  %``Initial indications for the production of a strongly coupled plasma in
  %Pb+Pb collisions at $\sqrt{s_{NN}} = 2.76$ TeV,''
  Phys.\ Rev.\  C {\bf 83}, 031901 (2011).
  %[arXiv:1011.6328 [nucl-ex]].
  %%CITATION = PHRVA,C83,031901;%%

\bibitem{Bozek:2010wt}
  P.~Bozek, M.~Chojnacki, W.~Florkowski and B.~Tomasik,
  %``Hydrodynamic predictions for Pb+Pb collisions at $\sqrt{S_{NN}}$ = 2.76
  %TeV,''
  Phys.\ Lett.\ {\bf B694}, 238 (2010);
  %[arXiv:1007.2294 [nucl-th]].
  %%CITATION = PHLTA,B694,238;%%
%\bibitem{Bozek:2011wa}
  P.~Bozek,
  %``Components of the elliptic flow in Pb-Pb collisions at 2.76 TeV,''
  %Phys.\ Lett.\  B 
  {\it ibid.} {\bf B699}, 283 (2011).
  %[arXiv:1101.1791 [nucl-th]].
  %%CITATION = PHLTA,B699,283;%%

\bibitem{Hirano:2010jg}
  T.~Hirano, P.~Huovinen and Y.~Nara,
  %``Elliptic flow in U+U collisions at sqrt{s_{NN}} = 200 GeV and in Pb+Pb
  %collisions at sqrt{s_{NN}} = 2.76 TeV: Prediction from a hybrid approach,''
  Phys.\ Rev.\  C {\bf 83}, 021902 (2011).
  %[arXiv:1010.6222 [nucl-th]].
  %%CITATION = PHRVA,C83,021902;%%

\bibitem{Schenke:2010rr}
  B.~Schenke, S.~Jeon and C.~Gale,
  %``Elliptic and triangular flow in event-by-event (3+1)D viscous 
  %  hydrodynamics,''
  Phys.\ Rev.\ Lett.\  {\bf 106}, 042301 (2011).
  %[arXiv:1009.3244 [hep-ph]].
  
\bibitem{Schenke:2011tv}
  B.~Schenke, S.~Jeon and C.~Gale,
  %``Anisotropic flow in sqrt(s)=2.76 TeV Pb+Pb collisions at the LHC,''
  Phys.\ Lett.\  {\bf B702}, 59 (2011).
  %[arXiv:1102.0575 [hep-ph]].
  %%CITATION = PHLTA,B702,59;%%

\bibitem{Song:2011qa}
  H.~Song, S.~A.~Bass and U.~Heinz,
  %``Elliptic flow in 200 A GeV Au+Au collisions and 2.76 A TeV Pb+Pb
  %collisions: insights from viscous hydrodynamics + hadron cascade hybrid
  %model,''
  Phys.\ Rev.\ C {\bf 83}, 054912 (2011).
  %[arXiv:1103.2380].
  %%CITATION = ARXIV:1103.2380;%%

\bibitem{Shen:2010uy}
  C.~Shen, U.~Heinz, P.~Huovinen and H.~Song,
  %``Systematic parameter study of hadron spectra and elliptic flow from
  %  viscous hydrodynamic simulations of Au+Au collisions at sqrt(s_NN) =
  %  200 GeV,''
  Phys.\ Rev.\ C {\bf 82}, 054904 (2010).
  %[arXiv:1010.1856 [nucl-th]].
  %%CITATION = ARXIV:1010.1856;%%

\bibitem{Shen:2011eg}
  C.~Shen, U.~Heinz, P.~Huovinen and H.~Song,
  %``Radial and elliptic flow in Pb+Pb collisions at the Large Hadron Collider from viscous hydrodynamic,''
  Phys.\ Rev.\ C {\bf 84}, 044903 (2011).
  %[arXiv:1105.3226 [nucl-th]].
  
\bibitem{Hirano:2005xf}
  T.~Hirano, U.~Heinz, D.~Kharzeev, R.~Lacey and Y.~Nara,
  %``Hadronic dissipative effects on elliptic flow in ultrarelativistic
  %heavy-ion collisions,''
  Phys.\ Lett.\ {\bf B636}, 299 (2006).
  %[arXiv:nucl-th/0511046].
  %%CITATION = PHLTA,B636,299;%%
  
\bibitem{Drescher:2006pi}
  A.~Adil, H.~J.~Drescher, A.~Dumitru, A.~Hayashigaki and Y.~Nara,
  %``The eccentricity in heavy-ion collisions from color glass condensate
  %initial conditions,''
  Phys.\ Rev.\  C {\bf 74}, 044905 (2006);
  %[arXiv:nucl-th/0605012].
  %%CITATION = PHRVA,C74,044905;%%
%\bibitem{Drescher:2007ax}
  H.~J.~Drescher and Y.~Nara,
  %``Eccentricity fluctuations from the Color Glass Condensate at RHIC and
  %LHC,''
  %Phys.\ Rev.\  C 
  {\it ibid.} {\bf 76}, 041903 (2007).
  %[arXiv:0707.0249 [nucl-th]].
  %%CITATION = PHRVA,C76,041903;%%
  
\bibitem{Hirano:2009ah}
  T.~Hirano and Y.~Nara,
  %``Eccentricity fluctuation effects on elliptic flow in relativistic heavy ion
  %collisions,''
  Phys.\ Rev.\  C {\bf 79}, 064904 (2009);
  %[arXiv:0904.4080 [nucl-th]].
  %%CITATION = PHRVA,C79,064904;%%
%\bibitem{Hirano:2009bd}
  %T.~Hirano and Y.~Nara,
  %``Eccentricity Fluctuation in Initial Conditions of Hydrodynamics,''
  and
  Nucl.\ Phys.\ {\bf A830}, 191C (2009).
  %[arXiv:0907.2966 [nucl-th]].
  %%CITATION = NUPHA,A830,191C;%%

\bibitem{Heinz:2009cv}
  U.~Heinz, J.~S.~Moreland and H.~Song,
  %``Viscosity from elliptic flow: The Path to precision,''
  Phys.\ Rev.\  {\bf C80}, 061901 (2009).
  %[arXiv:0908.2617 [nucl-th]].
  
\bibitem{Qiu:2011iv}
  Z.~Qiu and U.~Heinz,
  %``Event-by-event shape and flow fluctuations of relativistic heavy-ion
  %collision fireballs,''
  Phys.\ Rev.\  C {\bf 84}, 024911 (2011).
  %[arXiv:1104.0650 [nucl-th]].
  %%CITATION = PHRVA,C84,024911;%%

\bibitem{Alver:2010gr}
  B.~Alver and G.~Roland,
  %``Collision geometry fluctuations and triangular flow in heavy-ion 
  %  collisions,''
  Phys.\ Rev.\  C {\bf 81}, 054905 (2010).
  %[arXiv:1003.0194 [nucl-th]].

\bibitem{Adare:2011tg}
  A.~Adare {\it et al.}  (PHENIX Collaboration),
  %``Measurements of Higher-Order Flow Harmonics in Au+Au Collisions at
  %sqrt(s_NN) = 200 GeV,''
  arXiv:1105.3928 [nucl-ex];
  %%CITATION = ARXIV:1105.3928;%%
%\cite{Lacey:2011av}
%\bibitem{Lacey:2011av}
  R.~Lacey {\it et al} (PHENIX Collaboration),
  %``PHENIX Measurements of Higher-order Flow Harmonics in Au+Au collisions at Root_s = 200 GeV,''
  in {\it Quark Matter 2011}, {\it J. Phys.} G, in press [arXiv:1108.0457 [nucl-ex]].

\bibitem{Sorensen:2011fb}
  P.~Sorensen{\it et al.} (STAR Collaboration),
  %``Higher Flow Harmonics in Heavy Ion Collisions from STAR,''
  in {\it Quark Matter 2011}, {\it J. Phys.} G, in press [arXiv:1110.0737 [nucl-ex]].
  %%CITATION = ARXIV:1110.0737;%%

\bibitem{ALICE:2011vk}
  K. Aamodt {\it et al.} (ALICE Collaboration),
  %``Higher harmonic anisotropic flow measurements of charged particles in 
  %  Pb-Pb collisions at 2.76 TeV,''
  Phys.\ Rev.\ Lett.\  {\bf 107}, 032301 (2011);
%\bibitem{ALICE:2011yba}
  R.~Snellings {\it et al.} (ALICE Collaboration),
  %``Anisotropic flow at the LHC measured with the ALICE detector,''
  in {\it Quark Matter 2011}, {\it J. Phys.} G, in press [arXiv:1106.6284 [nucl-ex]];
  %%CITATION = ARXIV:1106.6284;%%
%\bibitem{Krzewicki:2011ee}
  M.~Krzewicki {\it et al.} (ALICE Collaboration), 
  %``Elliptic and triangular flow of identified particles at ALICE,''
  {\it ibid.} [arXiv:1107.0080 [nucl-ex]].

\bibitem{CMSflow}
  S. Chatrchyan {\it et al.} (CMS Collaboration),
  %``Measurement of higher-order harmonic flow in Pb+Pb collisions at center-of-mass energy = 2.76 TeV,"
  CERN preprint CMS-PAS-HIN-11-005;
  %
  J.~Velkovska {\it et al.} (CMS Collaboration), in {\it Quark Matter 2011},
  J.\ Phys.\ G, in press [http://cdsweb.cern.ch/record/1366652].
    
\bibitem{Steinberg:2011dj}
  P.~Steinberg {\it et al.} (ATLAS Collaboration),
  %``Recent Heavy Ion Results with the ATLAS Detector at the LHC,''
  J. Phys. G, in press [arXiv:1107.2182 [nucl-ex]];
%\bibitem{ATLASflow}
  J.~Jia {\it et al.}  (ATLAS Collaboration),
  %``Measurement of elliptic and higher order flow from ATLAS experiment at the LHC,''
  J.\ Phys.\ G, in press [arXiv:1107.1468 [nucl-ex]].

\bibitem{Alver:2010dn}
  B.~H.~Alver, C.~Gombeaud, M.~Luzum, and J.~Y.~Ollitrault,
  %``Triangular flow in hydrodynamics and transport theory,''
  Phys.\ Rev.\  C {\bf 82}, 034913 (2010).
  %[arXiv:1007.5469 [nucl-th]].
  %%CITATION = PHRVA,C82,034913;%%

\bibitem{Petersen:2010cw}
  H.~Petersen, G.-Y.~Qin, S.~A.~Bass and B.~M\"uller,
  %``Triangular flow in event-by-event ideal hydrodynamics in Au+Au collisions at $\sqrt{s_{\rm NN}}=200A$ GeV,''
  Phys.\ Rev.\  C {\bf 82}, 041901 (2010).
  %[arXiv:1008.0625 [nucl-th]].

\bibitem{Qin:2010pf}
  G.-Y.~Qin, H.~Petersen, S.~A.~Bass and B.~M\"uller,
  %``Translation of collision geometry fluctuations into momentum anisotropies in relativistic heavy-ion collisions,''
  Phys.\ Rev.\  C {\bf 82}, 064903 (2010).
  %[arXiv:1009.1847 [nucl-th]].
  
\bibitem{Luzum:2010sp}
  M.~Luzum,
  %``Collective flow and long-range correlations in relativistic heavy ion collisions,''
  Phys.\ Lett.\  {\bf B696}, 499-504 (2011).
  %[arXiv:1011.5773 [nucl-th]].
  
\bibitem{Xu:2011fe}
  J.~Xu and C.~M.~Ko,
  %``Triangular flow in heavy ion collisions in a multiphase transport model,''
  Phys.\ Rev.\  {\bf C84}, 014903 (2011).
  %[arXiv:1103.5187 [nucl-th]].

\bibitem{Luzum:2011mm}
  M.~Luzum,
  %``Flow fluctuations and long-range correlations: elliptic flow and beyond,''
   in {\it Quark Matter 2011}, {\it J. Phys.} G, in press [arXiv:1107.0592 [nucl-th]].
  
\bibitem{Chaudhuri:2011qm}
  A.~K.~Chaudhuri,
  %``Fluctuating initial conditions and fluctuations in elliptic and triangular flow,''
  arXiv:1108.5552 [nucl-th].

\bibitem{Schenke:2011bn}
  B.~Schenke, S.~Jeon, C.~Gale,
  %``Higher flow harmonics from (3+1)D event-by-event viscous hydrodynamics,''
  arXiv:1109.6289 [hep-ph].
  
\bibitem{Lacey:2010hw}
  R.~A.~Lacey, R.~Wei, N.~N.~Ajitanand, and A.~Taranenko,
  %``Initial eccentricity fluctuations and their relation to higher-order 
  %  flow harmonics,''
  Phys.\ Rev.\  C {\bf 83}, 044902 (2011);
  %[arXiv:1009.5230 [nucl-ex]].
  %%CITATION = PHRVA,C83,044902;%%
%\bibitem{Lacey:2011ug}
  R.~A.~Lacey, A.~Taranenko, N.~N.~Ajitanand and J.~M.~Alexander,
  %``Scaling of the higher-order flow harmonics: implications for
  %initial-eccentricity models and the 'viscous horizon',''
  arXiv:1105.3782 [nucl-ex].
  %%CITATION = ARXIV:1105.3782;%%
  
\bibitem{Shen:2011zc}
  C.~Shen {\it et al.},
  %, S.~A.~Bass, T.~Hirano, P.~Huovinen, Z.~Qiu, H.~Song and U.~Heinz,
  %``The QGP shear viscosity -- elusive goal or just around the corner?,''
   in {\it Quark Matter 2011}, {\it J. Phys.} G, in press
  [arXiv:1106.6350 [nucl-th]];

\bibitem{Qiu:2011fi}
  Z.~Qiu and U.~Heinz,
  %``Event-by-event hydrodynamics for heavy-ion collisions,''
  in {\it PANIC11}, AIP Conf. Proc., in press
  [arXiv:1108.1714 [nucl-th]].

\bibitem{Aamodt:2010pb}
  K.~Aamodt {\it et al.}  (ALICE Collaboration),
  %``Charged-particle multiplicity density at mid-rapidity in central Pb-Pb
  %collisions at sqrt(sNN) = 2.76 TeV,''
  Phys.\ Rev.\ Lett.\  {\bf 105}, 252301 (2010).
  %[arXiv:1011.3916 [nucl-ex]].
  %%CITATION = PRLTA,105,252301;%%

\bibitem{Aamodt:2010cz}
  K.~Aamodt {\it et al.}  (ALICE Collaboration),
  %``Centrality dependence of the charged-particle multiplicity density at
  %mid-rapidity in Pb-Pb collisions at sqrt(sNN) = 2.76 TeV,''
  Phys.\ Rev.\ Lett.\ {\bf 106}, 032301 (2011).
  %[arXiv:1012.1657 [nucl-ex]].
  %%CITATION = PRLTA,106,032301;%%

\bibitem{Heinz:2011kt}
  U.~Heinz, C.~Shen, and H.~Song,
  %``The viscosity of quark-gluon plasma at RHIC and the LHC,''
  in {\it PANIC11}, AIP Conf. Proc., in press
  [arXiv:1108.5323 [nucl-th]].
 
\bibitem{Qiu:future1}
  Z.~Qiu and U.~Heinz, manuscript in preparation.
 
\bibitem{Huovinen:2009yb}
  P.~Huovinen, P.~Petreczky,
  %``QCD Equation of State and Hadron Resonance Gas,''
  Nucl.\ Phys.\  {\bf A837}, 26 (2010).
  %[arXiv:0912.2541 [hep-ph]].
 
\bibitem{Teaney:2010vd}
  D.~Teaney and L.~Yan,
  %``Triangularity and Dipole Asymmetry in Heavy Ion Collisions,''
  Phys.\ Rev.\ C {\bf 83}, 064904 (2011).
  %[arXiv:1010.1876 [nucl-th]].
 
 \bibitem{Shuryak:2009cy}
  E.~Shuryak,
  %``The Cone, the Ridge and the Fate of the Initial State Fluctuations in Heavy Ion Collisions,''
  Phys.\ Rev.\  C {\bf 80}, 054908 (2009);
  %[arXiv:0903.3734 [nucl-th]].
%\bibitem{Staig:2010pn}
  P.~Staig and E.~Shuryak,
  %``The Fate of the Initial State Fluctuations in Heavy Ion Collisions. II The Fluctuations and Sounds,''
  arXiv:1008.3139 [nucl-th]
%\bibitem{Staig:2011wj}
  %P.~Staig, E.~Shuryak,
  %``The Fate of the Initial State Fluctuations in Heavy Ion Collisions. III The Second Act of Hydrodynamics,''
  and
  arXiv:1105.0676 [nucl-th]. 
 
\bibitem{Hirano:2002ds}
  T.~Hirano and K.~Tsuda,
  %``Collective flow and two pion correlations from a relativistic hydrodynamic model with early chemical freezeout,''
  Phys.\ Rev.\ C {\bf 66}, 054905 (2002);
  %[arXiv:nucl-th/0205043 [nucl-th]]. 
%\bibitem{Kolb:2002ve}
  P.~F.~Kolb, R.~Rapp,
  %``Transverse flow and hadrochemistry in Au+Au collisions at (S(NN))**(1/2) = 200-GeV,''
  {\it ibid.} 
  %Phys.\ Rev.\  
  C {\bf 67}, 044903 (2003);
  %[hep-ph/0210222].
%\bibitem{Hirano:2005wx}
  T.~Hirano and M.~Gyulassy,
  %``Perfect fluidity of the quark gluon plasma core as seen through its dissipative hadronic corona,''
  Nucl.\ Phys.\  {\bf A769}, 71 (2006);
  %[nucl-th/0506049].
%\bibitem{Huovinen:2007xh}
  P.~Huovinen,
  %``Chemical freeze-out temperature in hydrodynamical description of Au+Au collisions at s(NN)**(1/2) = 200-GeV,''
  Eur.\ Phys.\ J.\  {\bf A37}, 121 (2008).
  %[arXiv:0710.4379 [nucl-th]].
 
\bibitem{Aamodt:2010jd}
  K.~Aamodt  {\it et al.} (ALICE Collaboration),
  %``Suppression of Charged Particle Production at Large Transverse Momentum in Central Pb--Pb Collisions at $\sqrt{s_{NN}} = 2.76$ TeV,''
  Phys.\ Lett.\  {\bf B696}, 30 (2011).
  %[arXiv:1012.1004 [nucl-ex]]. 
 
\bibitem{Song:2008si}
  H.~Song and U.~Heinz,
  %``Multiplicity scaling in ideal and viscous hydrodynamics,''
  Phys.\ Rev.\ C {\bf 78}, 024902 (2008).
  %[arXiv:0805.1756 [nucl-th]].

\bibitem{Holopainen:2010gz}
  H.~Holopainen, H.~Niemi and K.~J.~Eskola,
  %``Event-by-event hydrodynamics and elliptic flow from fluctuating initial state,''
  Phys.\ Rev.\ C {\bf 83}, 034901 (2011).
  %[arXiv:1007.0368 [hep-ph]].
 
\bibitem{Miller:2003kd}
  M.~Miller and R.~Snellings,
  %``Eccentricity fluctuations and its possible effect on elliptic flow
  %measurements,''
  arXiv:nucl-ex/0312008.
  %%CITATION = NUCL-EX/0312008;%%
 
\bibitem{Voloshin:2007pc}
  S.~A.~Voloshin, A.~M.~Poskanzer, A.~Tang and G.~Wang,
  %``Elliptic flow in the Gaussian model of eccentricity fluctuations,''
  Phys.\ Lett.\  {\bf B659}, 537 (2008).
  %[arXiv:0708.0800 [nucl-th]].

\bibitem{Bhalerao:2006tp}
  R.~S.~Bhalerao and J.-Y.~Ollitrault,
  %``Eccentricity fluctuations and elliptic flow at RHIC,''
  Phys.\ Lett.\  {\bf B641}, 260 (2006).
  %[nucl-th/0607009].
 
\bibitem{Song:2010aq}
  H.~Song, S.~A.~Bass and U.~Heinz,
  %``Viscous QCD matter in a hybrid hydrodynamic+Boltzmann approach,''
  Phys.\ Rev.\ C {\bf 83}, 024912 (2011).
  %[arXiv:1012.0555 [nucl-th]].

\end{thebibliography}

\end{document}